\documentclass[prb,aps,showpacs,twocolumn,amsmath,amssymb,floatfix]{revtex4}
\pdfoutput=1
\usepackage{graphicx}% Include figure files
\usepackage{dcolumn}% Align table columns on decimal point
\usepackage{bm}% bold math

\begin{document}

\title{Three-electron anisotropic quantum dots in variable magnetic fields: 
exact results for excitation spectra, spin structures, and entanglement}

\author{Yuesong Li, Constantine Yannouleas, and Uzi Landman} 
\affiliation{
School of Physics, Georgia Institute of Technology, Atlanta, Georgia 30332-0430
}

\date{11 October 2007; Physical Review B, {\bf in press} }% It is always \today, today,

\begin{abstract}
Exact-diagonalization calculations for $N=3$ electrons in anisotropic quantum 
dots, covering a broad range of confinement anisotropies and strength of 
inter-electron repulsion, are presented for zero and low magnetic fields.
The excitation spectra are analyzed as a function of the strength of the 
magnetic field and for increasing quantum-dot anisotropy. 
Analysis of the intrinsic structure of the many-body wave functions through
spin-resolved two-point correlations reveals that the electrons 
tend to localize forming Wigner molecules. For certain ranges of dot parameters
(mainly at strong anisotropy), the Wigner molecules acquire a linear geometry, 
and the associated wave functions with a spin projection $S_z=1/2$ are similar 
to the representative class of strongly entangled states referred to as 
$W$-states. For other ranges of 
parameters (mainly at intermediate anisotropy), the Wigner molecules exhibit a 
more complex structure consisting of two mirror isosceles triangles. This
latter structure can be viewed as an embryonic unit of a zig-zag Wigner crystal
in quantum wires. The degree of entanglement in three-electron quantum dots can
be quantified through the use of the von Neumann entropy.
\end{abstract}

\maketitle

\section{Introduction}

Three-electron quantum dots are expected to attract a lot of attention in the 
near future due to several developments, both experimental and theoretical.
First, it was recently demonstrated\cite{elle06,marc04,kouw06} that 
detailed excitation spectra of two-electron quantum dots (in addition to 
earlier ground-state measurements\cite{kouw96,cior00}) can be measured,
and theoretically understood, as a function of the externally applied 
magnetic field. Thus, exploration of the 
excitation spectra of three-electron quantum dots appears to be a next step to
be taken. Second, three-qubit electron spin devices are expected to exhibit 
enhanced efficiency\cite{loss03,lida06,gorb06,vert07,loss07} for 
quantum-computing and quantum-information purposes compared to single-qubit 
and two-qubit ones. 

In this paper, we carry out exact diagonalization (EXD) studies for a 
three-electron single quantum dot under low and moderate magnetic fields. 
Unlike previous EXD studies\cite{hawr93,mikh02} that focused mainly on the 
ground states (GSs) of circular quantum dots,\cite{szaf06} 
we investigate, in addition, the 
excitation spectra for three electrons in quantum dots with a wide range of 
anisotropies. Moreover, consideration of anisotropic quantum dots allows us
to investigate the structure of the many-body wave functions with respect to 
strong-correlations effects, such as electron localization and formation of 
Wigner molecules with a linear or zig-zag geometry. 

Most importantly, we investigate here the feasibility of producing model 
quantum entangled states (i.e., the socalled $W$ 
states\cite{gorb06,cira00,woot00}), 
which are often employed in the mathematical treatment of quantum information 
and which have been experimentally realized with ultracold atoms in linear ion 
traps.\cite{roos04} We note that a main factor motivating our investigations is 
the different nature of the entangling agent, namely, the electromagnetic field 
in the case of heavy ions versus the two-body Coulomb interaction in the case of
electrons. 

We further mention other recent proposals in the context of solid 
state electronic devices for producing three-qubit entanglement. In particular, 
a scheme based on non-interacting electron-hole excitations in the Fermi sea was
investigated in Ref.\ \onlinecite{been04}. Unlike our present study that focuses
on the effect of the interparticle interaction, however, such 
{\it interaction-free\/} entanglement cannot\cite{been04} reproduce the 
symmetric $W$-state [see Eq.\ (\ref{wf3e3212}) below]. A different 
proposal\cite{vert07} for realizing interaction-free entanglement uses 
pair-correlation functions to study tripartite entanglement shared among the 
spins of three fermions in a Fermi gas. 

The exact diagonalization method that we use for the solution of the 
Schr\"{o}dinger equation corresponding to the Hamiltonian of three electrons 
interacting via a Coulomb potential in an anisotropic quantum dot, in 
conjunction with an analysis employing spin-resolved two-point correlation 
functions, allows us to gain deep insights into the nature of electronic states 
and three-qubit entanglement in real solid state devices. Additionally, the EXD 
method provides benchmark results, which could be used for assessment of the 
adequacy and relative accuracy of certain approximation schemes, including the 
model Heisenberg Hamiltonian for three localized electrons arranged in a ring 
geometry that was most recently used in an investigation of the entangled ground
states in a three-spin-qubit system.\cite{loss07}

\section{Outline of the exact diagonalization many-body method}
\label{exd}

We consider three electrons under zero or low magnetic field $(B)$ in a single
quantum dot. The corresponding many-body Hamiltonian is written as
\begin{equation}
{\cal H} = \sum_{i=1}^3 H(i) + 
\sum_{j>i=1}^3 \frac{e^2}{\kappa |{\bf r}_i -{\bf r}_j|},
\label{h3e}
\end{equation}
where $\kappa$ is the dielectric constant of the semiconductor material
(12.5 for GaAs). The single-particle Hamiltonian is given by
\begin{equation}
H=T+ V(x,y) + g^* \mu_B B\sigma,
\label{hone}
\end{equation}
where the last term is the Zeeman interaction, with $g^*$ being the effective 
Land\'{e} factor, $\mu_B$ the Bohr magneton, $B$ the perpendicular magnetic
field, and $\sigma=\pm 1/2$ the spin projection of an individual electron. 
The kinetic contribution in Eq.\ (\ref{hone}) is given by 
\begin{equation}
T=\frac{[{\bf p} - (e/c) {\bf A}({\bf r})]^2}{2m^*}, 
\label{hkin}
\end{equation}
with $m^*$ being the effective mass (0.067$m_e$ for GaAs) and   
the vector potential ${\bf A} ({\bf r}) = 0.5(-By\hat{\imath}+Bx\hat{\jmath})$
being taken according to the symmetric gauge. 
The external confining potential is denoted as $V(x,y)$, where
${\bf r} =x \hat{\imath} + y \hat{\jmath}$. 

The external potential is modeled by an anisotropic 2D oscillator
\begin{equation}
V(x,y) = \frac{1}{2} m^* (\omega_x^2 x^2+ \omega_y^2 y^2),
\label{vell}
\end{equation}
which reduces to a circular parabolic QD for $\omega_x=\omega_y=\omega_0$.
The ratio $\eta=\omega_x/\omega_y$ characterizes the degree of anisotropy
of the quantum dot, and it will be referred to thereafter as the anisotropy
parameter. Results will be presented for three cases: $\eta=1$ (circular),
$\eta=0.724$ (slightly anisotropic), and $\eta=1/2$ (strongly anisotropic).

We find the eigenstates of the many-body Hamiltonian (\ref{h3e}) using an
exact diagonalization method. Accordingly, we expand the many-body
wave function as a linear superposition,
\begin{equation}
\Psi^{\text{EXD}}({\bf r}_1, {\bf r}_2, {\bf r}_3) =
\sum_{1 \leq i < j < k \leq 2K} {\cal A}_{ijk} 
|\psi(1;i)\psi(2;j)\psi(3;k)\rangle,
\label{wf3e}
\end{equation}
where $|\psi(1;i)\psi(2;j)\psi(3;k)\rangle$ denotes a Slater determinant made
out of the three spin-orbitals $\psi(1;i)$, $\psi(2;j)$, and $\psi(3;k)$.
For the spin orbitals, we use the notation $\psi(1;i) = \varphi_i(1 \uparrow)$ 
if $1 \leq i \leq K$ and $\psi(1;i) = \varphi_{i-K}(1 \downarrow)$ if 
$K+1 \leq i \leq 2K$ [and similarly for $\psi(2;j)$ and $\psi(3;k)$].
$K$ is the maximum number of space orbitals $\varphi_i({\bf r})$ that are 
considered, with 
$\varphi_i(l \uparrow) \equiv \varphi_i({\bf r}_l) \alpha$ and 
$\varphi_i(l \downarrow) \equiv \varphi_i({\bf r}_l) \beta$ where $\alpha$ and
$\beta$ denote up and down spins, respectively. 
The space orbitals $\varphi_i({\bf r})$ are taken to coincide with the 
real eigenfunctions of a 2D anisotropic oscillator, that is, 
the index $i \equiv (m,n)$ and $\varphi_i ({\bf r}) = X_m(x) Y_n(y)$,
with $X_m(Y_n)$ being the eigenfunctions of the corresponding one-dimensional 
oscillators in the $x$($y$) direction with frequency $\omega_x$($\omega_y$). 
The parity operator ${\cal P}$ yields ${\cal P} X_m(x) = (-1)^m X_m(x)$, and 
similarly for $Y_n(y)$.
 
The total energies $E_{\text{EXD}}$ and the coefficients
${\cal A}_{ijk}$'s are obtained through a direct numerical diagonalization of
the matrix eigenvalue equation corresponding to the Hamiltonian in Eq.\
(\ref{h3e}). For the solution of this large scale, but sparse, matrix 
eigenvalue problem, we have used the ARPACK computer code.\cite{arpa}

%**************************** begin figure 1 ******************
\begin{figure}[t]
\centering\includegraphics[width=8.0cm]{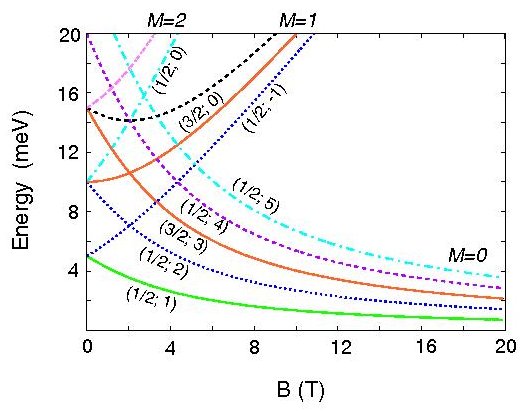}
\caption{(Color online)  %%%
Ground-state and excitation energy spectra 
[referenced to $3\hbar \sqrt{\omega_0^2 + \omega_c^2/4}$, with
$\omega_0= \sqrt{(\omega_x^2 + \omega_y^2)/2}$]
as a function of the magnetic field 
for $N=3$ non-interacting electrons in a circular quantum dot $(\eta=1)$. 
Parameters: external confinement $\hbar \omega_x = \hbar \omega_y =5$ meV;
dielectric constant $\kappa=\infty$; effective mass $m^*=0.067m_e$, 
effective Land\'{e} coefficient $g^*=0$. 
The labels $(S;L)$ denote the quantum numbers for the total spin and the
total angular momentum. Different Landau bands are denoted by the different
${\cal M}$ values.
The $S_z$ indices are not indicated, since the multiplets $(S=1/2,S_z)$ and
$(S=3/2,S_z)$ are degenerate in energy when $g^*=0$. 
}
\label{enspnint}
\end{figure}
%**************************** end figure 1 ******************

The EXD wave function (\ref{wf3e}) preserves by construction the
third projection $S_z$ of the total spin, since only Slater determinants with
a given $S_z$ value are used in the expansion. The exact diagonalization
automatically produces eigenfunctions of the square, ${\hat{\bf S}}^2$, of the 
total spin $\hat{\bf S} =\sum_{i=1}^3 {\hat{\sigma}}_i$. The corresponding
eigenvalues $S(S+1)$ are calculated with the help of the expression
\begin{equation}
\hat{{\bf S}}^2 |{\text{SD}} \rangle = 
\left [(N_\alpha - N_\beta)^2/4 + N/2 + \sum_{i<j} \varpi_{ij} \right ] 
|{\text{SD}}\rangle,
\end{equation}
where $|{\text{SD}}\rangle$ denotes a Slater determinant and the operator 
$\varpi_{ij}$ interchanges the spins of electrons $i$ and 
$j$ provided that their spins are different; $N_\alpha$ and $N_\beta$ denote 
the number of spin-up and spin-down electrons, respectively, while $N$ denotes
the total number of electrons.

Since the spin orbitals $\psi$'s are orthogonal, the Coulomb matrix elements 
between two Slater determinants are calculated using the Slater 
rules,\cite{szab} and the necessary two-body matrix elements between space 
orbitals 
\begin{equation}
\int \int d{\bf r}_1 d{\bf r}_2 
\varphi^*_i ({\bf r}_1) \varphi^*_j ({\bf r}_2) 
\frac{1}{ |{\bf r}_1 - {\bf r}_2 | }
\varphi_k ({\bf r}_1) \varphi_l ({\bf r}_2)
\label{vabgd}
\end{equation}
are calculated numerically. 
%via a center-of-mass transformation which eliminates
%the divergence at $|{\bf r}_1 - {\bf r}_2|=0$. 
We have found that this method produces numerically stable results in comparison
with algebraic expressions.\cite{vana}

%**************************** begin figure 2 ******************
\begin{figure}[b]
\centering\includegraphics[width=8.0cm]{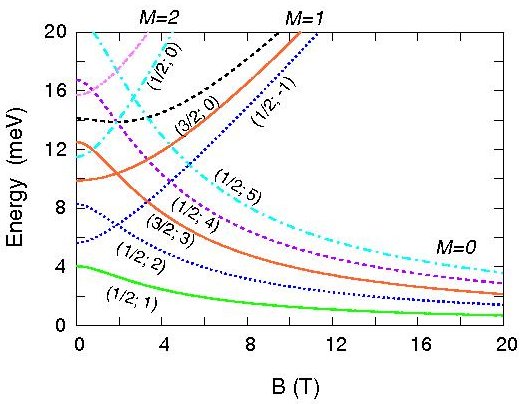}
\caption{(Color online)  %%%
Ground-state and excitation energy spectra
[referenced to $3\hbar \sqrt{\omega_0^2 + \omega_c^2/4}$, with
$\omega_0= \sqrt{(\omega_x^2 + \omega_y^2)/2}$]
as a function of the magnetic field for $N=3$ non-interacting electrons in a 
anisotropic quantum dot with anisotropy parameter $(\eta=0.724)$.
Parameters: external confinement $\hbar \omega_x = 4.23$ meV;
$\hbar \omega_y =5.84$ meV; dielectric constant $\kappa=\infty$; 
effective mass $m^*=0.070m_e$, effective Land\'{e} coefficient $g^*=0$.
The labels $(S;L)$ denote the quantum numbers for the total spin and the
total angular momentum in the corresponding circular quantum dot. Although
the total angular momentum is not a good quantum number for an anisotropic
quantum dot, we retain the labels $L$ here in order to facilitate 
comparison with the circular case in Fig.\ \ref{enspnint}.
The $S_z$ indices are not indicated, since the multiplets $(S=1/2,S_z)$ and
$(S=3/2,S_z)$ are degenerate in energy when $g^*=0$. 
}
\label{enspninteta}
\end{figure}
%**************************** end figure 2 ******************

\section{Energy spectra}

In this section, we study the ground-state and excitation spectra as a function
of an increasing magnetic field $B$ with an emphasis on the role of correlation
effects and the influence of the anisotropy.

%**************************** begin figure 3 ******************
\begin{figure}[t]
\centering\includegraphics[width=7.5cm]{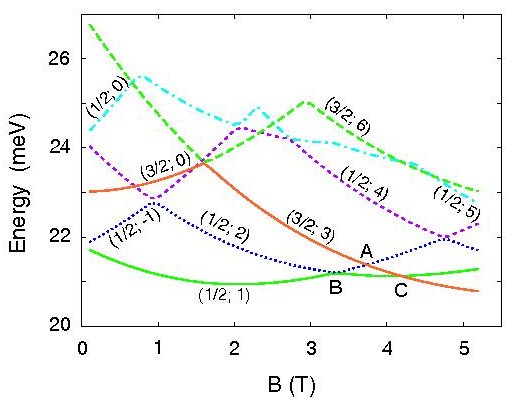}
\caption{(Color online)  %%%
Ground-state and excitation energy spectra 
[referenced to $3\hbar \sqrt{\omega_0^2 + \omega_c^2/4}$, with
$\omega_0= \sqrt{(\omega_x^2 + \omega_y^2)/2}$]
as a function of the magnetic field for $N=3$ interacting electrons in a 
circular quantum dot $(\eta=1)$. 
Parameters: external confinement $\hbar \omega_x = \hbar \omega_y = 5$ meV;
dielectric constant $\kappa=12.5$; effective mass $m^*=0.067m_e$;
effective Land\'{e} coefficient $g^*=0$. 
The labels $(S;L)$ denote the quantum numbers for the total spin and the
total angular momentum.
The $S_z$ indices are not indicated, since the multiplets $(S=1/2,S_z)$ and
$(S=3/2,S_z)$ are degenerate in energy when $g^*=0$. 
}
\label{enspcir}
\end{figure}
%**************************** end figure 3 ******************

To better understand the importance of correlations, we first display in Fig.\
\ref{enspnint} the spectra in the absence of the Coulomb interaction 
(non-interacting electrons) and for the case of a circular quantum dot. These 
energy spectra can be determined simply as 
$\sum_{i=1}^3 \epsilon_i^{\text{DF}}(B)$,
where $\epsilon_i^{\text{DF}}(B)$ are the Darwin-Fock energies for a single
electron.\cite{darw,fock,kouw01} The main trend is the formation of 
three-particle Landau bands (each with an
infinite number of states) that tend for $B \rightarrow \infty$ to the 
asymptotic energy levels $({\cal M}+3/2) \hbar \omega_c$, 
${\cal M}=0,1,2,\ldots$. Note that, for large magnetic fields 
($B \rightarrow \infty$), the reference energy  
$3\hbar \sqrt{\omega_0^2 + \omega_c^2/4}$, with
$\omega_0= \sqrt{(\omega_x^2 + \omega_y^2)/2}$], tends to
$3\hbar \omega_c /2$. In this limit, the states $(S,L)$, belonging to
the same Landau band ${\cal M}$, become degenerate in energy, converging to the
corresponding familiar Landau level (with index ${\cal M}$). 
Apart from an overall constant, the picture in Fig.\ \ref{enspnint} is the 
same as that found in the phenomenological ``constant-interaction'' 
model.\cite{kouw01} An important 
property is the absence of crossings between individual levels within each 
Landau band. A consequence of this is that the ground state 
at any $B$ has the same quantum numbers as the one at $B=0$, i.e., it has 
total spin $S=1/2$ and total angular momentum $L=1$.

The absence of crossings within each Landau band is a characteristic
property of non-interacting electrons, and it is independent of the anisotropy
of the external confinement. This point is illustrated in Fig.\ 
\ref{enspninteta} where the non-interacting three-electron spectra are
plotted for the case of a quantum-dot with moderate anisotropy ($\eta=0.724$).
[for the single-electron energies $\epsilon_i^{\text{DF}}(B)$ in an
elliptic quantum dot, see Ref.\ \onlinecite{chak94} and Ref.\ 
\onlinecite{kyri05}]. An inspection of Fig.\ \ref{enspninteta} shows that the 
anisotropy has an effect mainly for small values of the magnetic field
(by lifting the degeneracies at $B=0$). On the formation of the Landau bands 
at higher $B$, the anisotropy has practically no effect, and, in particular, 
it cannot induce level crossings within each Landau band).

\begin{figure}[b]
\centering\includegraphics[width=7.5cm]{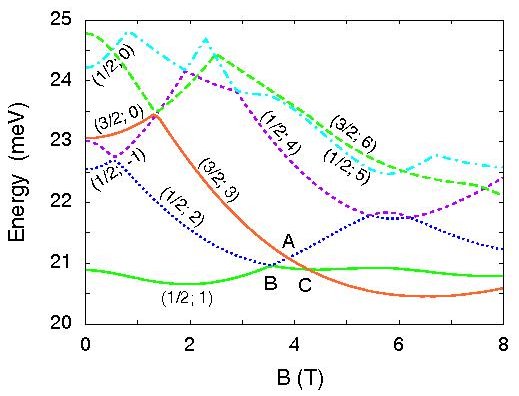}
\caption{(Color online)  %%%
Ground-state and excitation energy spectra 
[referenced to $3\hbar \sqrt{\omega_0^2 + \omega_c^2/4}$, with
$\omega_0= \sqrt{(\omega_x^2 + \omega_y^2)/2}$]
as a function of the magnetic field for $N=3$ interacting electrons in an 
elliptic quantum dot with intermediate anisotropy 
(anisotropy parameter $\eta=0.724$).
Parameters: external confinement $\hbar \omega_x = 4.23$ meV,
$\hbar \omega_y = 5.84$ meV; dielectric constant $\kappa=12.5$; effective mass
$m^*=0.070m_e$; effective Land\'{e} coefficient $g^*=0$. 
The labels $(S;L)$ denote the quantum numbers for the total spin and the
total angular momentum in the corresponding circular quantum dot.
The $S_z$ indices are not indicated, since the multiplets $(S=1/2,S_z)$ and
$(S=3/2,S_z)$ are degenerate in energy when $g^*=0$. 
Note the shrinking of the ABC triangle compared to the $\eta=1$ case 
shown in Fig.\ \ref{enspcir}.
}
\label{ensp0.7}
\end{figure}

\begin{figure}[t]
\centering\includegraphics[width=7.5cm]{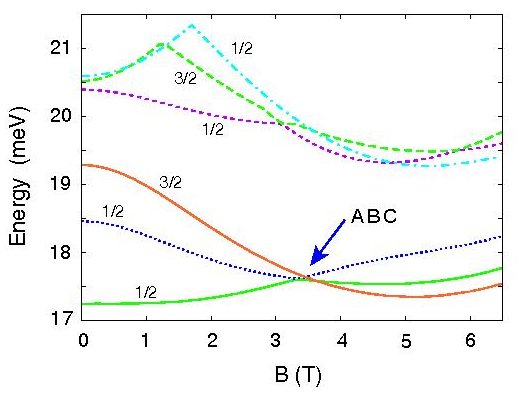}
\caption{(Color online)  %%%
Ground-state and excitation energy spectra 
[referenced to $3\hbar \sqrt{\omega_0^2 + \omega_c^2/4}$, with
$\omega_0= \sqrt{(\omega_x^2 + \omega_y^2)/2}$]
as a function of the magnetic field 
for $N=3$ electrons in an elliptic quantum dot with strong anisotropy 
(anisotropy parameter $\eta=1/2$).
Parameters: external confinement $\hbar \omega_x = 3.137$ meV,
$\hbar \omega_y = 6.274$ meV; dielectric constant $\kappa=12.5$; effective mass
$m^*=0.067m_e$; effective Land\'{e} coefficient $g^*=0$.
The single labels denote the quantum numbers for the total spin.
The $S_z$ indices are not indicated, since the multiplets $(S=1/2,S_z)$ and
$(S=3/2,S_z)$ are degenerate in energy when $g^*=0$. 
Note the collapse of the triangle ABC compared to the cases with 
$\eta=1$ (Fig.\ \ref{enspcir}) and $\eta=0.724$ (Fig.\ \ref{ensp0.7}), and
the appearance of a triple-point crossing.
}
\label{ensp0.5}
\end{figure}

Another property of the non-interacting spectra is the existence of several
degenerate levels (not shown in Figs.\ \ref{enspnint} and \ref{enspninteta})
associated with the excited states. We have searched for such degeneracies
by inducing a small lifting of them through the artificial use of a very weak
Coulomb repulsion specified by $\kappa=200$. For example, in the circular case
[see Fig.\ \ref{enspnint}], we found that the state $(1/2;2)$ is doubly 
degenerate, while the state $(3/2;3)$ is degenerate with two other $(1/2;3)$ 
states. These additional states move higher in energy as the strength of the
Coulomb interaction increases. We further found that the lifting of 
degeneracies is sufficiently strong for larger Coulomb repulsions with 
$\kappa \leq 12.5$ that all the curves in Figs.\ \ref{enspcir}, 
\ref{ensp0.7}, and \ref{ensp0.5} below are simple (i.e., the additional states 
present in the non-interacting case have been pushed much higher and fall 
outside the energy window shown). 

Turning on the interaction introduces correlation effects that lead to
important modifications of the non-interacting spectra shown in Figs.\
\ref{enspnint} and \ref{enspninteta}. Fig.\ \ref{enspcir} displays the
corresponding spectra for the same circular quantum dot as in Fig.\ 
\ref{enspnint}, but in the presence of a Coulomb repulsion 
with $\kappa=12.5$ (GaAs). Of course, a first effect
is the increase in the total energy, but the main difference 
from the non-interacting case in Fig.\ \ref{enspnint} is the presence of
crossings between levels within the same Landau band. As a result, within
the plotted range of magnetic fields, the ground-state total-spin quantum 
number remains $S=1/2$ at the first ground-state crossing (at point $B$), and 
then it changes to $S=3/2$ (at the second ground-state crossing at point
$C$). At the same time, the total angular momentum 
changes from $L=1$, to $L=2$, and then to $L=3$,
respectively. As long as the effective Land\'{e} coefficient $g^*=0$,
which is the case for the results presented in this section, this threefold 
alternation in the spin and angular momentum quantum numbers repeats itself 
ad-infinitum. We note that experimental observation of this threefold 
alternation may be forthcoming, since quantum dots with a vanishing Land\'{e} 
coefficient have been recently fabricated\cite{elle06} and were used already 
to measure two-electron excitation spectra.

The crossings of the curves associated with the three different pairs of 
quantum numbers $(S=1/2;L=1)$, $(1/2;2)$, and $(3/2;3)$ form a small triangle 
(labeled as ABC), which is located about $B \sim 3.4$ T. Anticipating the 
results for non-circular dots below, we note that this triangle tends 
to collapse to a single point with increasing anisotropy.

%**************************** begin figure 4 ******************
%**************************** end figure 4 ******************

Another prominent difference between the spectra of non-interacting (Fig.\ 
\ref{enspnint}) and interacting (Fig.\ \ref{enspcir}) electrons pertains 
to the degeneracies at $B=0$ between the $S=3/2$ and $S=1/2$ states 
that are lifted in the interacting-electrons case;
compare in particular the curves with quantum numbers $(1/2;2)$ and $(1/2;0)$ 
with the $(3/2;0)$ one. In contrast, the original degeneracies at $B=0$ of 
the $S=1/2$ states are unaffected by the interelectron interaction;
compare the curves $(1/2,1)$ and $(1/2,-1)$, as 
well as the ones labeled $(1/2,2)$ and $(1/2,0)$. However, these $S=1/2$ 
degenaracies at $B=0$ are lifted as a result of an increasing anisotropy of 
the quantum dot, as seen in Fig.\ \ref{ensp0.7}.

%**************************** begin figure 5 ******************
%**************************** end figure 5 ******************

Next, we explore the effect of increasing the anisotropy of the quantum dot. 
In particular, keeping the
same strength for the Coulomb interaction $(\kappa=12.5)$, we present two 
representative anisotropy cases, i.e., $\eta=0.724$ (intermediate anisotropy, 
see Fig.\ \ref{ensp0.7}), and $\eta=1/2$ (strong anisotropy closer to a 
quasilinear case, see Fig.\ \ref{ensp0.5}).

Inspection of the results for the case of intermediate anisotropy
(Fig.\ \ref{ensp0.7}), reveals that compared to Fig.\ \ref{enspcir} the 
spectra are distorted, but they maintain the overall topology of the 
circular dot. As a result, we have been able to use the same pairs of labels in
naming the different curves, even though the second label does not have
the meaning of an angular momentum (the total angular momentum is not conserved
for $\eta \neq 1$). There are two main differences from the circular case:
i) the degeneracies at $B=0$ between the $S=1/2$ states are lifted, and
ii) there is a marked rounding of all the $S=1/2$ curves in the beginning,
so that they do not intersect the vertical energy axis at sharp angles as is the 
case with Fig.\ \ref{enspcir}. This initial rounding and bending of the 
energy curves due to the anisotropy has been experimentally 
observed\cite{elle06,kyri02} in two-electron quantum dots.

In the case of strong anisotropy (Fig.\ \ref{ensp0.5}), the spectra have
evolved to such an extent that only little relation to the circular case can be
traced, and as a result we use a single label signifying the total spin to 
distinguish them. An important feature that emerges is that the three curves 
with lowest energies (two $S=1/2$ and one $S=3/2$ curve) 
form a band that is well 
separated from the other excited states. The existence of such an isolated 
lowest-energy band is important for validating simple two-qubit and 
three-qubit models introduced in quantum computation and quantum information 
theory.\cite{loss03,kyri07}

Another remarkable feature of the strong-anisotropy case is the appearance of a 
non-trivial triple-point crossing lying on the ground-state curve (see arrow in
Fig.\ \ref{ensp0.5}), which is created from the collapse of the ABC
triangle between the two $S=1/2$ and the one $S=3/2$ lowest-in-energy curves
(compare Figs.\ \ref{enspcir} and \ref{ensp0.7}). This low-energy non-trivial 
triple point [forming within the lowest Landau band (${\cal M}=0$)] is due to
the effect of the Coulomb interaction, and it is to be contrasted to other
trivial triple-point crossings at much higher energies arising from the
intersection of the lowest Landau band with the ${\cal M}=1$ and ${\cal M}=2$ 
higher Landau bands, and which are present even 
in the non-interacting limit [see, e.g., the 
triple crossing at (2.0 T; 14.2 meV) in Fig.\ \ref{enspnint}]. It would be of 
interest to analyze whether the recently observed\cite{aust07} triple-point 
crossings in deformed quantum dots are non-trivial or trivial in the sense 
described above. 

Before leaving this Section, we note that the spin multiplets $(S=1/2,S_z)$ 
and $(S=3/2,S_z)$ are 
degenerate in energy when $g^*=0$, which was the case for the energy spectra 
presented in Figs. 1-5. At a given magnetic field, this degeneracy is naturally
lifted when $g^* \neq 0$; however, the final total energies can be easily 
calculated by adding the Zeeman term $g^*\mu_B B S_z$ to the spectral curves 
displayed in these figures. Furthermore, for a given pair $(S,S_z)$, the Zeeman 
term does not influence the intrinsic structure of the many-body EXD wave 
function [i.e., the expansion over constituent Slater determinants, see 
Eq.\ (\ref{wf3e})], and thus taking $g^*=0$ does not effect the results for 
electron densities, conditional probability distributions, and von Neumann 
entropies presented below.

%**************************** begin figure 6 ******************
\begin{figure}[t]
\centering{\includegraphics[width=8cm]{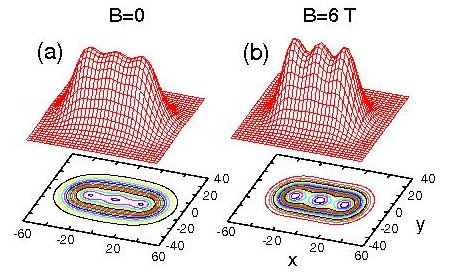}}
\caption{(Color online) %%%
Exact-diagonalization electron densities for the ground states of $N=3$ 
electrons in an anisotropic dot with parameters $\hbar\omega_x=3.137$ meV, 
$\hbar\omega_y=6.274$ meV $(\eta=1/2)$, effective mass $m^*=0.067m_e$, 
dielectric constant $\kappa=12.5$ (GaAs).
(a): The case of zero magnetic field, $B=0$. (b) The case with
a magnetic field $B=6$ T.
Lengths in nm. The electron densities are in arbitrary units, but with
the same scale in both panels.
}
\label{den12k125}
\end{figure}
%**************************** end figure 6 ******************

\section{Many-body wave functions for strong anisotropy ($\eta=1/2$)} 
\label{seceta12}

\subsection{$S=1/2$ ground states: Evolution of electron densities as a
function of the inter-electron repulsion}

\begin{figure}[t]
\centering{\includegraphics[width=8cm]{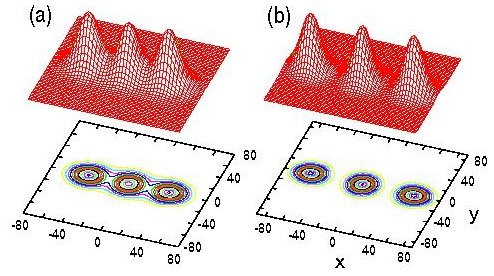}}
\caption{(Color online) %%%
Exact-diagonalization electron densities at zero magnetic field ($B=0$) for 
the ground state of $N=3$ electrons in an anisotropic dot with parameters 
$\hbar\omega_x=3.137$ meV, $\hbar\omega_y=6.274$ meV $(\eta=1/2)$,
$m^*=0.067m_e$. (a): dielectric constant $\kappa=3.0$. 
(b): dielectric constant $\kappa=1.0$.
Lengths in nm. The electron densities are in arbitrary units, but with
the same scale as in Fig.\ \ref{den12k125} for both panels.
}
\label{den12k3k1}
\end{figure}

When the confining potential lacks circular symmetry, charge localization
is reflected directly in the single-particle electron densities. Indeed,
electron localization is visible in Figs.\ \ref{den12k125} and  
\ref{den12k3k1}, which display the electron densities for $N=3$ electrons
in an anisotropic quantum dot with $\eta=1/2$. Fig.\ \ref{den12k125}
illustrates the evolution of electron localization with increasing magnetic
field in the case of a weaker Coulomb repulsion $(\kappa=12.5)$. One sees that
already at $B=0$, the electron density is shaped linearly for all practical 
purposes. However, the three peaks of the localized electrons are rather weak, 
which contrasts with the case of $B=6$ T [Fig.\ \ref{den12k125}(b)], where
the three electron peaks are sharply defined.

Fig.\ \ref{den12k3k1} [in conjunction with Fig.\ \ref{den12k125}(a)] 
illustrates the strengthening of electron localization as a function of
increasing Coulomb repulsion, i.e., decreasing dielectric constant $\kappa$, 
from a value of 12.5 [Fig.\ \ref{den12k125}(a)] to $\kappa=3$ 
[Fig.\ \ref{den12k3k1}(a)] and then to $\kappa=1$ [Fig.\ \ref{den12k3k1}(b)]. 
In this last case [Fig.\ \ref{den12k3k1}(b)], the three electrons are almost 
fully localized, with orbitals that exhibit practically zero mutual overlap.

Since we keep the average frequency, 
$\omega_0=\sqrt{(\omega_x^2 + \omega_x^2)/2}$, approximately constant
(i.e., $\hbar \omega_0 \approx 5.0$ meV) for all anisotropy cases studied
in this paper, decreasing the dielectric constant is equivalent to   
increasing the Wigner parameter\cite{yann99} $R_W$. At zero magnetic field, 
$R_W$ is widely used as a universal parameter to indicate the strength of
correlations, since it provides the relative strength of the Coulomb 
repulsion with respect to the quantum kinetic energy, i.e.,
\begin{equation}
R_W = \frac{e^2/(\kappa l_0)} {\hbar \omega_0},
\label{wigpar} 
\end{equation}
with the characteristic length $l_0=\sqrt{\hbar/(m^* \omega_0)}$. 
For the numerical values of $R_W$ 
associated with the cases studied here, see Fig.\ \ref{enteta12k}.

%**************************** begin figure 7 ******************
%**************************** end figure 7 ******************

\subsection{$S=1/2$ ground state: Spin resolved intrinsic structure for
strong repulsion ($\kappa=1$)}
\label{seceta12s12}

In the previous section, we saw that already the electron densities provide
partial information about the formation of a linear Wigner molecule within
an elliptic quantum dot. Indeed, from the
charge distributions in Figs.\ \ref{den12k125} and \ref{den12k3k1}, one 
can infer that the electrons are localized in three separate positions 
${\bf R}_1$, ${\bf R}_2$, and ${\bf R}_3$. If the electrons were spinless, this
situation could be approximately reproduced by a single Slater determinant
denoted as $|\bigcirc\bigcirc\bigcirc\rangle$. However, to probe the
spin distribution of the electrons, the exact-diagonalization charge densities 
do not suffice; one needs to consider spin-resolved two-point correlation
functions, defined as
%\begin{equation}
\begin{eqnarray}
&& \hspace{-0.8cm} P_{\sigma\sigma_0}({\bf r}, {\bf r}_0)=  \nonumber \\
&& \hspace{-0.5cm} \langle \Psi^{\text{EXD}} |
\sum_{i \neq j} \delta({\bf r} - {\bf r}_i) \delta({\bf r}_0 - {\bf r}_j)
\delta_{\sigma \sigma_i} \delta_{\sigma_0 \sigma_j}
|\Psi^{\text{EXD}}\rangle,
\label{sponcpd}
\end{eqnarray}
%\end{equation}
with the EXD many-body wave function given by equation (\ref{wf3e}).

Using a normalization constant
\begin{equation}
{\cal N}(\sigma,\sigma_0,{\bf r}_0) = 
\int P_{\sigma\sigma_0}({\bf r}, {\bf r}_0) d{\bf r},
\label{norm}
\end{equation}
we further define a related conditional probability distribution (CPD) as
\begin{equation}
{\cal P}_{\sigma\sigma_0}({\bf r}, {\bf r}_0) =
P_{\sigma\sigma_0}({\bf r}, {\bf r}_0)/{\cal N}(\sigma,\sigma_0,{\bf r}_0),
\label{cpd}
\end{equation}
having the property 
$\int {\cal P}_{\sigma\sigma_0}({\bf r}, {\bf r}_0) d{\bf r} =1$.
The spin-resolved CPD gives the spatial probability distribution of 
finding the 
remaining electrons with spin projection $\sigma$ under the condition that the 
third electron is located (fixed) at ${\bf r}_0$ with spin projection 
$\sigma_0$; $\sigma$ and $\sigma_0$ can be either up $(\uparrow$) or
down ($\downarrow$). 

Before examining such CPDs, evaluated for numerically determined EXD wave 
functions, it is instructive to consider on a qualitative 
level the spin structure of the wave functions that 
can be formed from three localized spin-orbitals only. In particular, we focus 
on the case with a total spin projection $S_z=1/2$, when the most general
three-orbital wave function is given by the superposition of three 
Slater determinants, i.e., by the expression 
\begin{equation} 
\Phi(S_z=\mbox{$\frac12$})=
a |\uparrow \downarrow  
\uparrow \;\rangle +
b |\uparrow \uparrow 
\downarrow \;\rangle +
c |\downarrow \uparrow 
\uparrow \;\rangle,
\label{wfspabc}
\end{equation}
with the normalization $a^2 + b^2 + c^2=1$. Unlike the circles used earlier
to indicate spinless electrons, the arrows in Eq.\ (\ref{wfspabc}) indicate 
the spin projections of the individual spin-orbitals.

%**************************** begin figure 8 ******************
\begin{figure}[t]
\centering{%%%%%%
\includegraphics[width=8.3cm]{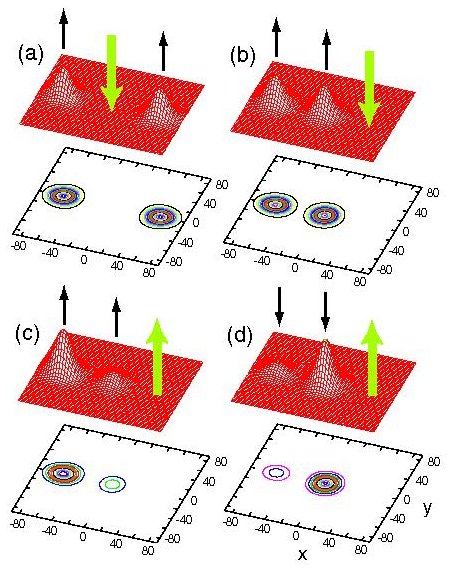}}
\caption{(Color online) %%%
Spin-resolved conditional probability distributions for the (1/2,1/2) ground 
state of $N=3$ electrons in an anisotropic dot at zero magnetic field 
$(B=0)$ with parameters $\hbar\omega_x=3.137$ meV, $\hbar\omega_y=6.274$ meV 
$(\eta=1/2)$, $m^*=0.067m_e$ and $\kappa=1$ [for the corresponding electron
density, see Fig.\ \ref{den12k3k1}(b)].
The heavy arrow (green online) indicates the location of the fixed electron
at ${\bf r}_0$ [see Eq.\ (\ref{cpd})], with the indicated spin projection
$\sigma_0$, i.e., up ($\uparrow$) or down ($\downarrow$).
(a) $\uparrow \downarrow$ CPD with the fixed spin-down electron located at the 
center.  
(b) $\uparrow \downarrow$ CPD with the fixed spin-down electron located on the 
right. 
(c) $\uparrow \uparrow$ CPD with the fixed spin-up electron located on the 
right. 
(d) $\downarrow \uparrow$ CPD with the fixed spin-up electron located on the 
right. 
The spin of the fixed electron is denoted by a thick arrow (green online).
Lengths in nanometers. The vertical axes are in arbitrary units, but the
scale is the same for all four panels.
}
\label{cpdk1}
\end{figure}
%**************************** end figure 8 ******************

The general states (\ref{wfspabc}) are a superposition of three Slater 
determinants and have attracted a lot of attention in the mathematical theory
of entanglement. Indeed, they represent a prototypical class of three-qubit
entangled states known as $W$-states.\cite{woot00} For general coefficients
$a$, $b$, and $c$, the states (\ref{wfspabc}) are not 
eigenfunctions of the square of total spin ${\hat{\bf S}}^2$
(while the exact-diagonalization wave functions in Eq.\ (\ref{wf3e}
are always good eigenfunctions of ${\hat{\bf S}}^2$). 
However, the special values of these coefficients that lead to good total-spin
quantum numbers can be determined.\cite{lida06,pauncz} In particular, using the
notation $\Phi (S,S_z;i)$ (where the index $i$ is employed in case of a 
degeneracy), one has
\begin{equation}
\sqrt{3} \Phi (\mbox{$\frac32$},\mbox{$\frac12$}) = 
| \uparrow
 \downarrow
 \uparrow \;\rangle +
| \uparrow
 \uparrow
 \downarrow \;\rangle+
| \downarrow
 \uparrow
 \uparrow \;\rangle 
\label{wf3e3212}
\end{equation}
(i.e., $a=b=c=1/\sqrt{3}$),
\begin{equation}
\sqrt{6} \Phi (\mbox{$\frac12$},\mbox{$\frac12$};1) =
2 | \uparrow
 \downarrow
 \uparrow \;\rangle 
- | \uparrow
 \uparrow
 \downarrow \;\rangle
- | \downarrow
 \uparrow
 \uparrow \;\rangle 
\label{wf3e12121}
\end{equation}
(i.e., $a=2/\sqrt{6}$, $b=c=-1/\sqrt{6}$), 
\begin{equation}
\sqrt{2}\Phi (\mbox{$\frac12$},\mbox{$\frac12$};2) =
| \uparrow
 \uparrow
 \downarrow \;\rangle
-| \downarrow
 \uparrow
 \uparrow \;\rangle 
\label{wf3e12122}
\end{equation}
(i.e., $a=0$, $b=1/\sqrt{2}$, $c=-1/\sqrt{2}$).

For completeness, we list the case for three fully spin-polarized localized
electrons (which of course is not a $W$-state).
\begin{equation}
\Phi (\mbox{$\frac32$},\mbox{$\frac32$}) = 
| \uparrow
 \uparrow
 \uparrow \;\rangle.
\label{wf3e3232}
\end{equation}
The wave functions with projections $S_z=-1/2$ and $S_z=-3/2$ are similar to 
the above, but with inverted single-particle spins.

Before proceeding further, we note that the term $W$-state is some times 
reserved for the symmetric form $\Phi (\mbox{$\frac32$},\mbox{$\frac12$})$ in
Eq.\ (\ref{wf3e3212}). This symmetric $W$-state has been experimentally
realized in linear ion traps.\cite{roos04} As we show below, quantum dots
offer the means for generating in addition the less symmetric forms
given by Eqs.\ (\ref{wf3e12121}) and (\ref{wf3e12122}). Such nonsymmetric
three-qubit states are sometime denoted\cite{cao03} as $W^\prime$-states (with a 
prime). In this paper, we do not make use of this distinction, and we refer to
both symmetric and nonsymmetric forms simply as $W$-states.

%**************************** begin figure 9 ******************
\begin{figure}[t]
\centering{%%%%%%
\includegraphics[width=8.3cm]{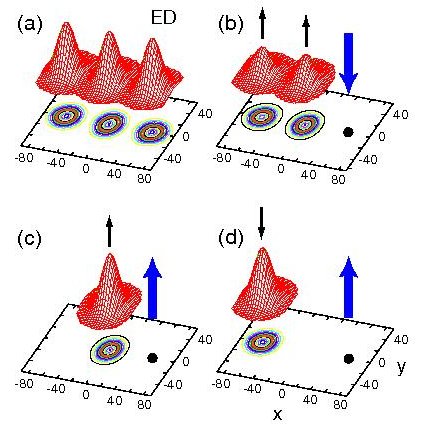}}
\caption{(Color online) %%%
Spin-resolved conditional probability distributions for the (1/2,1/2) first
excited state of $N=3$ electrons in an anisotropic dot at zero magnetic field
$(B=0)$ with parameters $\hbar\omega_x=3.137$ meV, $\hbar\omega_y=6.274$ meV
$(\eta=1/2)$, $m^*=0.067m_e$ and $\kappa=1$.
(a) electron density (ED).
(b) $\uparrow \downarrow$ CPD with the fixed spin-down electron located on the
right at (61,0).
(c) $\uparrow \uparrow$ CPD with the fixed spin-up electron located on the
right at (61,0).
(d) $\downarrow \uparrow$ CPD with the fixed spin-up electron located at the
right at (61,0).
The spin of the fixed electron is denoted by a thick arrow (blue online).
Lengths in nanometers. The vertical axes are in arbitrary units, but the
scale is the same for all four panels.
}
\label{cpdk1exci12}
\end{figure}
%**************************** end figure 9 ******************

In Fig.\ \ref{cpdk1}, we present several spin-resolved CPDs associated with
the EXD ground state at $B=0$ and strong anisotropy $\eta=1/2$, which is a
$\Psi^{\text{EXD}}(1/2,1/2)$ state [see Fig.\ \ref{ensp0.5}]. Although the
EXD expansion in Eq.\ (\ref{wf3e}) consists of a large number of Slater
determinants built from delocalized harmonic-oscillator orbitals, the CPD 
patterns in Fig.\ \ref{cpdk1} reveal an intrinsic structure similar to that
of the wave function $\Phi (\mbox{$\frac12$},\mbox{$\frac12$};1)$ in Eq.\
(\ref{wf3e12121}), which is made out of only three localized spin-orbitals.
In particular, when one requires that the fixed electron has a down spin and is
located at the center of the quantum dot, the spin-up electrons are
located on the left and right with equal weights [Fig.\ \ref{cpdk1}(a)].
Keeping the down spin-direction, but moving the fixed electron to the
right, reveals that the spin-up electrons are located on the left and
the center with equal weights [Fig.\ \ref{cpdk1}(b)]. Considering a spin-up
direction for the fixed electron and placing it on the right reveals that
the remaining spin-up electron is distributed on the left and the center of
the quantum dot with unequal weights; approximately 4 (left) to 1 (center) 
following the square of the coefficients in front of the determinants
$| \uparrow 
 \downarrow
 \uparrow \;\rangle$ ($a=2/\sqrt{6}$) and
$| \downarrow
 \uparrow
 \uparrow \;\rangle$ ($c=1/\sqrt{6}$) in the wave
function $\Phi (\mbox{$\frac12$},\mbox{$\frac12$};1)$ 
[see Eq.\ (\ref{wf3e12121})]. 
Similarly, considering a spin-up direction for the fixed electron and placing it
on the right reveals that the spin-down electron is distributed on the
left and the center of the quantum dot with unequal weights -- approximately 
1 (left) to 4 (center), in agreement with the weights of the Slater 
determinants in Eq.\ (\ref{wf3e12121}). 

\subsection{$S=1/2$ first excited state: Spin resolved intrinsic structure for
strong repulsion ($\kappa=1$)}
\label{seceta12s12exci}

In section \ref{seceta12s12}, we investigated the intrinsic structure of
the ground-state three-electron wave functions with total spin $S=1/2$ and for
the case of a strong anisotropy $\eta =1/2$. In this section, we analyze a
case of the first-excited EXD wave function with total spin $S=1/2$ and for 
the same strong anisotropy $\eta=1/2$, again at $B=0$ T and for strong 
interelectron repulsion $\kappa=1$. We denote this state as
$\Psi^{\text{EXD}}(1/2,1/2;2)$.

%**************************** begin figure 10 ******************
\begin{figure}[t]
\centering{%%%%%%
\includegraphics[width=8.3cm]{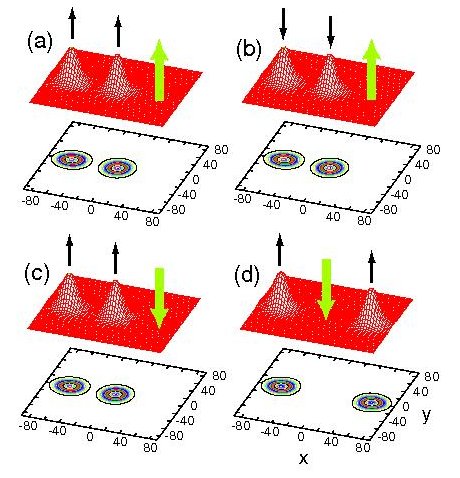}}
\caption{(Color online) %%%
Spin-resolved conditional probability distributions for the (3/2,1/2) second 
excited state of $N=3$ electrons in an anisotropic dot at zero magnetic field 
$(B=0)$ with parameters $\hbar\omega_x=3.137$ meV, $\hbar\omega_y=6.274$ meV 
$(\eta=1/2)$, $m^*=0.067m_e$ and $\kappa=1$.
(a) $\uparrow \uparrow$ CPD with the fixed spin-up electron located on the
right at (70,0).
(b) $\downarrow \uparrow $ CPD with the fixed spin-up electron located on the 
right at (70,0).
(c) $\uparrow \downarrow$ CPD with the fixed spin-down electron located on the 
right at (70,0).
(d) $\uparrow \downarrow$ CPD with the fixed spin-down electron located at the 
center.
The spin of the fixed electron is denoted by a thick arrow (green online).
Lengths in nanometers. The vertical axes are in arbitrary units, but the
scale is the same for all four panels.
}
\label{cpdk1exci32}
\end{figure}
%**************************** end figure 10 ******************

In Fig.\ \ref{cpdk1exci12}(a), we display the electron density (ED) for this 
second $S=1/2$ state, while in Figs.\ \ref{cpdk1exci12}(b,c,d), we display 
spin-resolved CPDs for the same state. From the charge density, we conclude 
that the three electrons form a sharply defined linear Wigner molecule.
The spin-resolved CPD with a spin-down fixed electron placed on the right 
[see Fig.\ \ref{cpdk1exci12}(b)] is similar to that in Fig.\ \ref{cpdk1}(b).
However, the two spin-resolved CPDs with a {\it spin-up\/} fixed electron 
placed on the right [see Figs.\ \ref{cpdk1exci12}(c) and \ref{cpdk1exci12}(d)]
are quite different from the corresponding CPDs in Figs.\ \ref{cpdk1}(c) and
\ref{cpdk1}(d). In fact, in both cases, only one single hump appears to the 
left of the fixed electron, located at the center for the remaining spin-up
electrons [Fig.\ \ref{cpdk1exci12}(c)], or on the left for the remaining 
spin-down electrons [Fig.\ \ref{cpdk1exci12}(d)].

This indicates that the intrinsic structure of the 
$\Psi^{\text{EXD}}(1/2,1/2;2)$ wave function
is close to that of $\Phi (\mbox{$\frac12$},\mbox{$\frac12$};2)$ in
Eq.\ (\ref{wf3e12122}), with $a=0$ and $b=-c$.

%**************************** begin figure 11 ******************
\begin{figure}[t]
\centering{%%%%%%
\includegraphics[width=8.3cm]{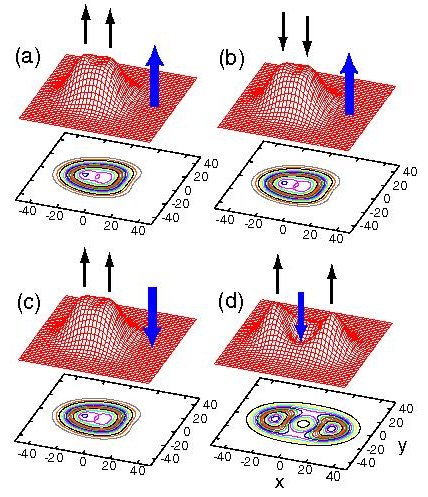}}
\caption{(Color online) %%%
Spin-resolved conditional probability distributions for the (3/2,1/2) ground 
state of $N=3$ electrons in an anisotropic dot at $B=5$ T with parameters 
$\hbar\omega_x=4.23$ meV, $\hbar\omega_y=5.84$ meV 
$(\eta=0.724)$, $m^*=0.070m_e$ and $\kappa=12.5$.
(a) $\uparrow \uparrow$ CPD with the fixed spin-up electron located on the
right at (30,0).
(b) $\downarrow \uparrow$ CPD with the fixed spin-up electron located on the 
right at (30,0).
(c) $\uparrow \downarrow$ CPD with the fixed spin-down electron located on the 
right at (30,0).
(d) $\uparrow \downarrow$ CPD with the fixed spin-down electron located at the 
center.
The spin of the fixed electron is denoted by a thick arrow (blue online).
Lengths in nanometers. The vertical axes are in arbitrary units, but the
scale is the same for all four panels.
}
\label{cpdk125eta0.7}
\end{figure}
%**************************** end figure 11 ******************

\subsection{$S=3/2$ second excited state: Spin resolved intrinsic structure for
strong repulsion ($\kappa=1$)}

In sections \ref{seceta12s12} and \ref{seceta12s12exci}, 
we investigated the intrinsic structure of 
the many-body three-electron wave functions with total spin $S=1/2$ and for
the case of a strong anisotropy $\eta =1/2$. In this section, we analyze a 
case of an EXD wave function with total spin $S=3/2$ and for the same strong  
anisotropy $\eta=1/2$, again at $B=0$ T. In particular, we analyze the 
intrinsic structure of a $\Psi^{\text{EXD}}(3/2,1/2)$ wave function that is 
the second excited state for these parameters.

In Fig.\ \ref{cpdk1exci32}, we display spin-resolved CPDs for this $S=3/2$ 
excited state. A remarkable feature is that for a fixed electron placed on the
right all three CPDS, $\uparrow \uparrow$ [Fig.\ \ref{cpdk1exci32}(a)],
$\downarrow \uparrow$ [Fig.\ \ref{cpdk1exci32}(b)], and 
$\uparrow \downarrow$ [Fig.\ \ref{cpdk1exci32}(c)] coincide. This indicates
that the intrinsic structure of the $\Psi^{\text{EXD}}(3/2,1/2)$ wave function
is close to that of $\Phi (\mbox{$\frac32$},\mbox{$\frac12$})$ in
Eq.\ (\ref{wf3e3212}), with all three coefficients equal to each other,
i.e., $a=b=c$.

%**************************** begin figure 12 ******************
\begin{figure}[t]
\centering{%%%%%%
\includegraphics[width=8.3cm]{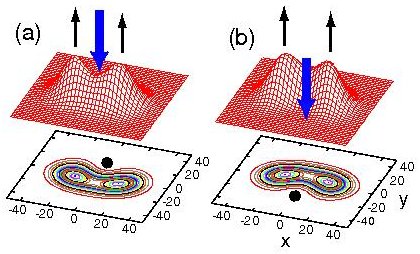}}
\caption{(Color online) %%%
Spin-resolved conditional probability distributions for the (3/2,1/2) ground 
state of $N=3$ electrons in an anisotropic dot at $B=5$ T with parameters 
$\hbar\omega_x=4.23$ meV, $\hbar\omega_y=5.84$ meV 
$(\eta=0.724)$, $m^*=0.070m_e$ and $\kappa=12.5$.
(a) $\uparrow \downarrow$ CPD with the fixed spin-down electron located on the 
$y$-axis at (0,20) (solid dot).
(b) $\uparrow \downarrow$ CPD with the fixed spin-down electron located on the 
$y$-axis at (0,-20) (solid dot).
The spin of the fixed electron is denoted by a thick arrow (blue online).
Lengths in nanometers. The vertical axes are in arbitrary units, but the
scale is the same for all panels in this figure and in Fig.\ 
\ref{cpdk125eta0.7}.
}
\label{cpdk125eta0.7down}
\end{figure}
%**************************** end figure 12 ******************

Taking into account the $\uparrow \downarrow$ CPD with the fixed electron at
the center of the quantum dot, it is clear that the geometric arrangement of 
the three localized electrons is linear. Arrangements that are more 
complicated than the linear 
geometry can emerge, however, for a range of different 
parameters, as is discussed in section \ref{sec0.7} below. 

\section{Many-body wave functions for intermediate anisotropy ($\eta=0.724$)}

\subsection{Moderate repulsion ($\kappa=12.5$)}
\label{sec0.7}

In this section, we analyze a 
case of an EXD wave function with total spin $S=3/2$ and for the intermediate 
anisotropy $\eta=0.724$. In particular, we analyze the intrinsic structure of 
a $\Psi^{\text{EXD}}(3/2,1/2)$ wave function that is the ground state at a
magnetic field $B=5$ T (see Fig.\ \ref{ensp0.7}). 

In Fig.\ \ref{cpdk125eta0.7}, we display spin-resolved CPDs for this
ground state. A remarkable feature is that for a fixed electron placed on the
right all three CPDS, $\uparrow \uparrow$ [Fig.\ \ref{cpdk125eta0.7}(a)],
$\downarrow \uparrow$ [Fig.\ \ref{cpdk125eta0.7}(b)], and 
$\uparrow \downarrow$ [Fig.\ \ref{cpdk125eta0.7}(c)] coincide. This indicates
that the intrinsic structure of the $\Psi^{\text{EXD}}(3/2,1/2)$ wave function
is close to that of $\Phi (\mbox{$\frac32$},\mbox{$\frac12$})$ in
Eq.\ (\ref{wf3e3212}), with all three coefficients equal to each other, 
$a=b=c$.

However, these CPDs, as well as the  $\uparrow \downarrow$ CPD with the fixed 
spin-down electron at the center [Fig.\ \ref{cpdk125eta0.7}(d)], are broader
along the $y$-direction compared to the CPDs associated with the linear 
molecular arrangement in Fig.\ \ref{cpdk1exci32}. This suggests that, for an 
intermediate anisotropy ($\eta=0.724$), the intrinsic structure of 
$\Psi^{\text{EXD}}(3/2,1/2)$ is 
more complicated. Indeed, as demonstrated in Fig.\ \ref{cpdk125eta0.7down}
where the fixed spin-down electron is successively placed away from the 
$x$-axis at (0, 20 nm) and at (0, -20 nm), the intrinsic structure corresponds
to a superposition of two molecular isomers, each one described by a
three-orbital wave function  $\Phi (\mbox{$\frac32$},\mbox{$\frac12$})$, but
with the three localized spin-orbitals located on the vertices of two 
isosceles triangles, each one being a mirror reflection (relative to the 
$x$-axis) of the other. The base of the first isosceles triangle lies at
-6 nm [Fig.\ \ref{cpdk125eta0.7down}(a)] and that of the second one at 6 nm
[Fig.\ \ref{cpdk125eta0.7down}(a)] off the $x$-axis (in the $y$-direction).

%**************************** begin figure 13 ******************
\begin{figure}[t]
\centering{%%%%%%
\includegraphics[width=8.3cm]{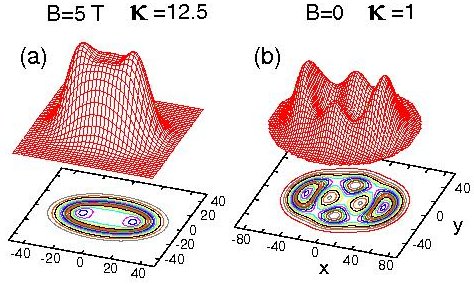}}
\caption{(Color online) %%%
Exact-diagonalization electron densities for the ground state 
of $N=3$ electrons
in an anisotropic quantum dot with parameters $\hbar\omega_x=4.23$ meV, 
$\hbar\omega_y=5.84$ meV ($\eta=0.724$, intermediate anisotropy) and 
$m^*=0.070m_e$.
(a) the (3/2,1/2) ground state at $B=5$ T and $\kappa=12.5$.
(b) the (1/2,1/2) ground state at $B=0$ and $\kappa=1$ (strong interelectron
repulsion).
Lengths in nm. The electron densities are in arbitrary units, with
a different scale in each panel.
}
\label{den0.7k125}
\end{figure}
%**************************** end figure 13 ******************

The two-triangle configuration discussed for three electrons above may be seen 
as the embryonic precursor of a quasilinear structure of two intertwined 
``zig-zag'' crystalline chains. Such {\it double\/} zig-zag 
crystalline chains may also be related to the {\it single\/} zig-zag 
Wigner-crystal chains discussed recently in relation to spontaneous spin 
polarization in quantum wires.\cite{klir06,piac04}

It is interesting to inquire of how this two-triangle structure is reflected
in the spatial distribution of the electron densities. 
Indeed, in Fig.\ \ref{den0.7k125}(a), we display
the electron density associated with the (3/2,1/2) ground state at $B=5$ T.
We note in particular the absence of a third peak at the center of the quantum
dot. Instead, two rather small peaks appear at (0,20 nm) and (0,-20 nm),
in agreement with the two-triangle internal structure revealed by the
CPD analysis. 

%**************************** begin figure 14 ******************
\begin{figure}[t]
\centering{%%%%%%
\includegraphics[width=8.3cm]{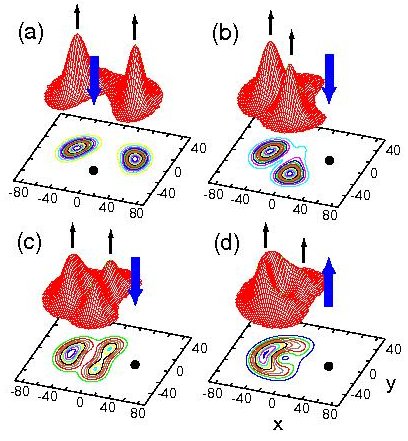}}
\caption{(Color online) %%%
Spin-resolved conditional probability distributions for the (1/2,1/2) ground
state of $N=3$ electrons in an anisotropic dot at $B=0$ with parameters
$\hbar\omega_x=4.23$ meV, $\hbar\omega_y=5.84$ meV
$(\eta=0.724)$, $m^*=0.070m_e$ and $\kappa=1$.
(a) $\uparrow \downarrow$ CPD with the fixed spin-down electron located on the
$y$-axis at (0,-20) (solid dot).
(b) $\uparrow \downarrow$ CPD with the fixed spin-down electron located off 
center at (40,11) (solid dot).
(c) $\uparrow \downarrow$ CPD with the fixed spin-down electron located on the
$x$-axis at (43,0) (solid dot).
(d) $\uparrow \uparrow$ CPD with the fixed spin-ip electron located on the
$x$-axis at (43,0) (solid dot).
The spin of the fixed electron is denoted by a thick arrow (blue online).
Lengths in nanometers. The vertical axes are in arbitrary units, but the
scale is the same for all panels in this figure.
}
\label{cpdk1eta0.7}
\end{figure}
%**************************** end figure 14 ******************

\subsection{Strong repulsion ($\kappa=1$)}

We further display in Fig.\ \ref{den0.7k125}(b) the corresponding electron
density for the (1/2,1/2) ground state at $B=0$ and for a strong Coulomb
repulsion $(\kappa=1)$ at the intermediate anisotropy $\eta=0.724$. As a 
result of the enhanced electron localization, the electron density 
exhibits pronounced peaks whose locations form a clearly defined diamond;
this indicates again the presence of a two-triangle internal 
configuration.\cite{note24} 
The detailed interlocking of the two triangular configurations is further
revealed in the spin-resolved CPDs that are displayed in Fig.\ 
\ref{cpdk1eta0.7}. From the CPDs in Figs.\  \ref{cpdk1eta0.7}(a) and
\ref{cpdk1eta0.7}(b), it can be concluded that one triangle is formed by the 
points ${\bf R}_1 \approx (0,-20)$ nm, ${\bf R}_2 \approx (-43,10)$ nm, and
${\bf R}_3 \approx (43,10)$ nm, while the second one (its mirror) is formed by
the points ${\bf R}_1^\prime \approx (0,20)$ nm, 
${\bf R}_2^\prime \approx (-43,-10)$ nm, and
${\bf R}_3^\prime \approx (43,-10)$ nm. The $\uparrow \downarrow$ [Fig.\ 
\ref{cpdk1eta0.7}(c)] and $\uparrow \uparrow$ [Fig.\ \ref{cpdk1eta0.7}(d)]
CPDS with the fixed electron on the right at (43,0) nm are similar to those
in Figs.\ \ref{cpdk1}(b) and \ref{cpdk1}(c), respectively, with the 
difference that the central hump is clearly a double one. 
This indicates that each triangular configuration is associated with a wave 
function of the form $\Phi (\mbox{$\frac12$},\mbox{$\frac12$};1)$ given in
Eq.\ (\ref{wf3e12121}).

Naturally, the regime of a linear configuration versus a two-triangle one 
depends on both the strength of the interaction and the anisotropy. Detailed 
studies of the phase boundary between these two intrinsic structures are, 
however, left for a future investigation. 

\section{Degree of entanglement}

The many-body wave functions for $N=3$ electrons analyzed 
in the previous sections
are highly entangled states, since they cannot be reduced to a single Slater 
determinant. For special ranges of the dot parameters, we showed that they
acquire the same internal structure as the prototypical $W$-states. In this 
section, we demonstrate that the degree of entanglement can be further 
quantified through the use of the von Neumann entropy ${\cal S}_{\text{vN}}$
for indistinguishable fermions which (in analogy to the two-electron case 
\cite{you01,busc06,ihn07,yann07}) is defined  as
\begin{equation}
{\cal S}_{\text{vN}} = - {\text{Tr}} (\rho \log_2 \rho) + {\cal C},
\label{entvn}
\end{equation}
where ${\cal C}$ is a constant (see below for choosing its value) and the 
single-particle density matrix is given by
\begin{equation}
\rho_{\nu\mu} = \frac{\langle \Psi^{\text{EXD}} | a^\dagger_\mu a_\nu |
\Psi^{\text{EXD}} \rangle}
{\sum_\mu \langle \Psi^{\text{EXD}} | a^\dagger_\mu a_\mu |
\Psi^{\text{EXD}} \rangle}, 
\label{densmat}
\end{equation}
and is normalized to unity, i.e.,
${\text{Tr}} \rho=1$. The Greek indices $\mu$ (or $\nu$) count the spin 
orbitals $\psi({\bf r};\mu)$ that span the single-particle space (of dimension
$2K$; see Section \ref{exd}).
Note that, in keeping with previous literature on two 
electrons,\cite{you01,ihn07,yann07} the logarithms are taken to base two.

Naturally, for calculating numerically the matrix elements $\rho_{\nu\mu}$, we
use the expansion (\ref{wf3e}) to get
\begin{equation}
\langle \Psi^{\text{EXD}} | a^\dagger_\mu a_\nu |
\Psi^{\text{EXD}} \rangle =
\sum_{I,J} {\cal A}^*_I {\cal A}_J 
\langle {\text{SD}}(I) | a^\dagger_\mu a_\nu | {\text{SD}} (J) \rangle,
\label{sdij}
\end{equation}
where the following conventions for the indices $I$ (or $J$) apply:
$I \rightarrow (ijk)$ and $|{\text{SD}}(I) \rangle = 
|\psi(1;i)\psi(2;j)\psi(3;k)\rangle$. The matrix elements 
$\langle {\text{SD}}(I) | a^\dagger_\mu a_\nu | {\text{SD}} (J) \rangle$ 
between Slater determinants that enter in Eq.\ (\ref{sdij}) simply equal
$\pm 1$ or vanish. The single-particle density $\rho$ in Eq.\ (\ref{densmat}) is
in general non-diagonal. Thus we further perform numerically a diagonalization
of $\rho$, and we use the new diagonal elements $\tilde{\rho}_{\mu\mu}$ to 
straightforwardly calculate the von Neumann entropy in Eq.\ (\ref{entvn}).

\begin{figure}[t]
\centering\includegraphics[width=7.5cm]{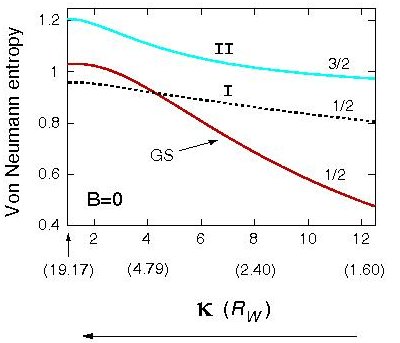}
\caption{(Color online)  %%%
Von Neumann entropy at zero magnetic field for the three lowest EXD states with 
$S_z=1/2$ as a function of the dielectric constant $\kappa$ [or equivalently the
Wigner parameter $R_W$; see Eq.\ (\ref{wigpar})] for $N=3$ electrons in an 
anisotropic quantum dot with strong anisotropy (anisotropy parameter $\eta=1/2$).
Parameters: external confinement $\hbar \omega_x = 3.137$ meV,
$\hbar \omega_y = 6.274$ meV; effective mass $m^*=0.067m_e$.
The single labels $3/2$ and $1/2$ denote the quantum numbers for the total spin.
The ground state (GS), and the first (I) and second (II)
excited states are indicated. The horizontal arrow indicates the direction
of increasing correlations. 
According to our convention, the von Neumann entropy for a single
determinant vanishes. 
Although the energy gaps between the three
EXD states diminish with decreasing $\kappa$ (they are quasidegenerate for
$\kappa=1$), the relative energy ordering remains unchanged in the plotted
range. 
}
\label{enteta12k}
\end{figure}

\begin{figure}[t]
\centering\includegraphics[width=7.5cm]{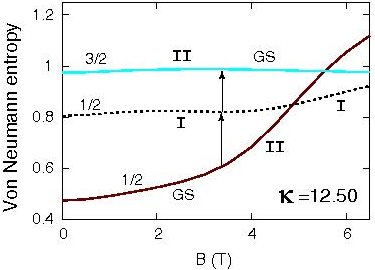}
\caption{(Color online)  %%%
Von Neumann entropy for the three lowest EXD states with $S_z=1/2$ as a 
function of the magnetic field for $N=3$ electrons in an anisotropic quantum 
dot with strong anisotropy (anisotropy parameter $\eta=1/2$).
Parameters: external confinement $\hbar \omega_x = 3.137$ meV,
$\hbar \omega_y = 6.274$ meV; dielectric constant $\kappa=12.5$; effective mass
$m^*=0.067m_e$.
The single labels $3/2$ and $1/2$ denote the quantum numbers for the total spin.
The vertical arrows indicate the discontinuous jump in the von Neumann entropy 
of the ground state at $B=3.4$ T, where the ground-state quantum numbers change 
character, first from ($1/2,1/2;1$) to ($1/2,1/2;2$), and then immediately to 
($3/2,1/2$). 
The ground state (GS), and the first (I) and second (II) excited states are 
indicated, both to the left and to the right of the vertical arrows. 
According to our convention, the von Neumann entropy for a single
determinant vanishes.
}
\label{enteta12}
\end{figure}

As was discussed in Refs.\ \onlinecite{busc06,ihn07,yann07}, 
the von Neumann entropy provides a natural measure of entanglement in the
case of interacting indistinguishable fermions. In this case, the entanglement 
is related\cite{ecke02} to quantum correlations that are intrinsic to the 
many-body wave function, i.e., 
${\cal S}_{\text{vN}}$ quantifies the fact that strongly correlated states 
comprise a larger number of significant Slater determinants compared to weakly 
correlated ones.
Accordingly, one expects that ${\cal S}_{\text{vN}}$ increases when the many-body
correlations increase (i.e., when $R_W$ and $B$ increase). This was the case
indeed for the $N=2$ quantum dot,\cite{ihn07,yann07} but we have
found that it also holds true for the $N=3$ quantum dot, as can be seen from
Fig.\ \ref{enteta12k} and Fig.\ \ref{enteta12}.

In the case when the many-body wave function reduces to a single Slater 
determinant, i.e., when the expansion coefficients reduce to 
${\cal A}_I = \delta_{I, I_0}$, all the matrix elements
$\rho_{\mu\nu}$ vanish except three diagonal ones (corresponding to three
fully occupied spin orbitals) which are equal to 1/3; then
$-{\text{Tr}}(\rho \log_2 \rho)=\log_2 3=1.5850$. Since the entanglement due
to the Pauli exchange principle by itself cannot be used as a resource for 
quantum-information processing,\cite{busc06,ghir04} we take the constant 
${\cal C}$ in Eq.\ (\ref{entvn}) to be 
\begin{equation}
{\cal C}=-\log_2 N,
\label{const}
\end{equation}
and as a result the von Neumann entropy for a single Slater determinant
vanishes in our convention. 

%**************************** begin figure 15 ******************
%**************************** end figure 15 ******************

In Fig.\ \ref{enteta12k}, we plot the von Neumann entropy for the three lowest 
EXD states with $S_z=1/2$ as a function of $\kappa$ ($R_W$) for $N=3$ 
electrons in an anisotropic quantum dot for a strong anisotropy with anisotropy 
parameter $\eta=1/2$. It is apparent that the von Neumann entropy increases
for all three states as $R_W$ increases 
($\kappa$ decreases) and the electrons become more localized. 

At $\kappa=12.50$ (corresponding to weaker correlations), the von Neumann
entropies for the three states are clearly non-vanishing, indicating that
these EXD states are far from being close to a single Slater determinant. 
On the other hand, it is natural to expect that the EXD states will reduce to
single Slater determinants at the non-interacting limit. To check this
expectation, we have carried out an EXD calculation for the same QD parameters 
described in the caption of Fig.\ \ref{enteta12k}, but with a very large
$\kappa=10000$ in order to approximately mimick the non-interacting limit. In 
this latter case, we found that indeed the ground state 
[with $(S=1/2; S_z=1/2)$ is practically a 
single Slater determinant made out from the three spin orbitals 
$(m=0,n=0; \uparrow)$, $(m=0,n=0; \downarrow)$, and $(m=1,n=0; \uparrow)$ 
[the lowest-in-energy spatial orbital $(m=0,n=0)$ being doubly occupied; see 
Section \ref{exd} for the meaning of indices $m$ and $n$). 
We also found that the corresponding $S_{\text{vN}}$ is practically zero.

However, due to the $\eta=1/2$ anisotropy, one has 
$2\hbar\omega_x=\hbar \omega_y$, 
which gives rise to a high degree of energy degeneracy among excited Slater 
determinants with good total spin. For example, the Slater determinant 
$|(m=0,n=0; \uparrow), (m=1,n=0; \downarrow), (m=1,n=0; \uparrow) \rangle$
is degenerate to the determinant 
$|(m=0,n=0; \uparrow), (m=0,n=0; \downarrow), (m=0,n=1; \uparrow) \rangle$.
In this situation, a small $e-e$ interaction is sufficient to produce strong
mixing of the degenerate Slater determinants, and as a result the corresponding
$S_{\text{vN}}$ values for excited states were found to be non-vanishing. 
These findings are
reflected in Fig.\ \ref{enteta12k} where, for $\kappa=12.50$ (weakest Coulomb
repulsion in the plotted range), the $S_{\text{vN}}$ value for the EXD
ground state is noticeably lower than the values for the two excited states.

As was demonstrated in Section \ref{seceta12}, at zero magnetic field 
and strong Coulomb repulsion (e.g., $\kappa=1$), 
the three electrons are well separated and localized, and 
their EXD wave functions are equivalent to the forms given in Eq.\ 
(\ref{wf3e12121}) (GS), Eq.\ (\ref{wf3e12122}) (first excited state, I), and 
Eq.\ (\ref{wf3e3212}) (second excited state, II). These forms are special cases
of the general form in Eq.\ (\ref{wfspabc}) for which another measure of
entanglement, called the tangle and specifying the reduced tripartite 
entanglement among the three localized spin qubits,\cite{woot00} 
can be applied. 

The tangle can be calculated\cite{woot00} from the coefficients a, b, and c, and
it was found that it vanishes for all cases covered by the general form in
Eq.\ (\ref{wfspabc}). In this respect, the von Neumann entropy for three well 
separated electrons studied here exhibits qualitatively a very different
behavior, since the values of $S_{\text{vN}}$ at $\kappa=1$ are all different,
as seen from Fig.\ \ref{enteta12k}. In particular, we note that in this 
case the EXD value of $S_{\text{vN}}$ for the I state is lower than that of the
GS state; this naturally reflects the fact that the first excited EXD state in 
this limit is effectively composed of only two Slater determinants 
[see Eq.\ (\ref{wf3e12122})] compared to the three Slater determinants 
associated [see Eq.\ (\ref{wf3e12121})] with the EXD ground state. 

%**************************** begin figure 16 ******************
%**************************** end figure 16 ******************

In Fig.\ \ref{enteta12}, we plot the von Neumann entropy for the three lowest 
EXD states with $S_z=1/2$ as a function of the magnetic field for $N=3$ 
electrons in an anisotropic quantum dot with the same parameters as those 
for the energy spectra in Fig.\ \ref{ensp0.5} (strong anisotropy with anisotropy 
parameter $\eta=1/2$, as also was the case with Fig.\ \ref{enteta12k}). 
It is apparent that the von Neumann entropy increases
for all three states as the magnetic field increases and the electrons
become more localized. An interesting feature is the discontinuous jump 
(around $B=3.4$ T) in the von Neumann entropy of the EXD ground state.
This jump is illustrated by the vertical arrows and is associated with
the triple ABC point in Fig.\ \ref{ensp0.5}. This discontinuity in
the ground-state $S_{\text{vN}}$ arises from the sudden change in the intrinsic 
structure (in term of constituent Slater determinants) of the ground state, 
as the latter changes its quantum numbers first from ($1/2,1/2;1$) to 
($1/2,1/2;2$), and then again immediately to ($3/2,1/2$) at the triple point.

\section{Summary}

We have presented extensive exact-diagonalization calculations for $N=3$
electrons in anisotropic quantum dots, and for a broad range of
anisotropies and strength of inter-electron repulsion. We have analyzed the
excitation spectra both as a function of the magnetic field and as a function
of increasing anisotropy. A main finding was the appearance of triple-crossing
points in the ground-state energy curves for stronger anisotropies.

Analysis of the intrinsic structure of the many-body wave functions through
spin-resolved conditional probability distributions revealed that for all 
examined cases (including those with parameters corresponding to currently
fabricated quantum dots) the electrons localize forming Wigner molecules. For 
certain ranges of dot parameters (mainly at strong anisotropy), the Wigner 
molecules acquire a linear geometry, and the associated wave functions with 
a spin projection $S_z=1/2$ are similar to the socalled $W$-states that are a 
prototype of entangled states. For other ranges of parameters (mainly at 
intermediate anisotropy), the Wigner molecules exhibit a more complex structure 
consisting of two mirror isosceles triangles. This latter structures can be 
considered as an embryonic unit of a zig-zag Wigner crystal in quantum wires.

Finally, we demonstrated that the degree of entanglement in three-electron
quantum dots can be quantified via the von Neumann entropy, in analogy
with studies on two-electron quantum dots.

\end{document}